\documentclass[showpacs,aps,prb,reprint,superscriptaddress,preprintnumbers]{revtex4-2}

\usepackage{amsmath}
\usepackage{amssymb}
\usepackage{graphicx}
\usepackage{dcolumn}
\usepackage[dvipsnames]{xcolor}
\usepackage{amsthm,amssymb}
\usepackage{mathrsfs}
\usepackage{color}
\usepackage{endnotes}
\usepackage{bm}
\usepackage{epsfig}
\usepackage{array}
\usepackage{ulem}

\begin{document}

\title[]{Point-contact Andreev reflection 
spectroscopy of layered superconductors with device-integrated diamond anvil cells}

\author{Che-hsuan~Ku$^\S$}%
\affiliation{Department of Physics, The Chinese University of Hong Kong, Shatin N.T., Hong Kong, China}%
\author{Omargeldi Atanov$^\S$}%
\affiliation{Department of Physics, The Hong Kong University of Science and Technology, Clear Water Bay, Kowloon, Hong Kong, China}%
\author{King Yau Yip}%
\author{Wenyan Wang}%
\author{Siu Tung Lam}%
\affiliation{Department of Physics, The Chinese University of Hong Kong, Shatin N.T., Hong Kong, China}%
\author{Jiayu Zeng}%
\affiliation{Department of Physics, The Hong Kong University of Science and Technology, Clear Water Bay, Kowloon, Hong Kong, China}%
\author{Wei Zhang}%
\author{Zheyu Wang}%
\author{Lingfei Wang}%
\author{Tsz Fung Poon}%
\affiliation{Department of Physics, The Chinese University of Hong Kong, Shatin N.T., Hong Kong, China}%
\author{Rolf Lortz}%
 \email[]{lortz@ust.hk}
\affiliation{Department of Physics, The Hong Kong University of Science and Technology, Clear Water Bay, Kowloon, Hong Kong, China}%
\author{Swee K. Goh}
 \email[]{skgoh@cuhk.edu.hk}
\affiliation{Department of Physics, The Chinese University of Hong Kong, Shatin N.T., Hong Kong, China}%

\date{\today}

\begin{abstract}
{
Superconductors that can be mechanically exfoliated are an interesting platform for exploring superconducting properties tuned by layer thickness. These layered superconductors are also expected to exhibit sensitivity to applied pressure. While pressure has been demonstrated to be an effective way of tuning bulk superconductors, analogous studies on superconducting thin flakes have been limited due to technical challenges. In particular, spectroscopic measurements under pressure remain insufficiently explored. In this work, we functionalized the diamond anvil cell technique for point-contact Andreev reflection spectroscopy (PCAR) measurement on thin-flake materials under pressure, offering the opportunity to obtain spectroscopic information on superconductivity. To validate the feasibility of this method, we have conducted PCAR measurements on iron-selenide thin flakes to extract temperature-dependent superconducting gap values under ambient and high pressure. Combine with the proven magnetotransport capability, our method provides a conceptually simple tool for a detailed examination of thin-flake superconductors under pressure.
}
\end{abstract}

\maketitle

\section{Introduction}
Layered materials are of considerable interest in condensed matter physics because their unique crystal structures, in which the bonding along the interplanar direction is significantly weakened, can host exotic quantum phases such as unconventional superconductivity or density-wave states~\cite{Qiu2021,Saito2016,Wang2023}. These intriguing quantum phases can be tuned by various parameters, such as hydrostatic pressure, thickness reduction or magnetic field~\cite{Sun2016,Xie2021}. In particular, hydrostatic pressure is an interesting tuning parameter because it decreases interatomic distances and consequently modifies the electronic hopping integral, giving rise to a different electronic structure which may promote new phenomena. Superconductivity has been successfully induced or enhanced by pressure in a range of layered materials~\cite{Pan2015,Kang2015,Alireza2008,Qi2016,Takahashi2017,Lee2018,Heikes2018,Hu2019,Hu2020,Xie2021,Sun2016}. For instance, in the transition metal dichalcogenide (TMD) WTe$_{2}$, superconductivity emerges only under high pressure~\cite{Pan2015,Kang2015}. Moreover, in the kagome metal CsV$_{3}$Sb$_{5}$ thin flake, the superconducting transition temperature $T_c$ is enhanced from 4.3~K at ambient pressure to 7.8~K at $\sim$20~kbar, which is a nearly two-fold enhancement~\cite{wang2024,Zhang2024,Zhang2023}, while a three-fold enhancement has been reported in bulk CsV$_3$Sb$_5$~\cite{chen2021,yu2021}. For the iron-based superconductor iron-selenide (FeSe), which we will discuss in this work, $T_c$ varies in a double-dome manner under pressure, and a $T_c$ as high as $\sim$40~K is achieved~\cite{Mizuguchi2008,Masaki2009}. Meanwhile, the spin density wave order can also be induced under pressure, in the region of the phase diagram where $T_c$ is maximized~\cite{Margadonna2009,Sun2016,Xie2021,Terashima2016}. Therefore, pressure serves as a powerful tool for investigating exotic superconductivity in layered materials.

The complex Cooper pairing mechanisms in layered superconductors pose the need for investigating the superconducting gaps, thus driving the development of spectroscopic measurement techniques. While angle-resolved photoemission spectroscopy (ARPES) and scanning tunneling microscopy (STM) are reliable tools for probing the superconducting gap, the demands of the surface morphology and the high-vacuum environment hinder their application under high pressure. On the contrary, point-contact Andreev reflection spectroscopy (PCAR) is a `wire-based' technique, and the high pressure community possesses the know-how to bring wires into the high-pressure region. For PCAR, the regime of transport should ideally be ballistic which is achieved by making the metal-superconductor contact smaller than the electron mean free path. In that case, the energy of the electrons can be fully determined, allowing the implementation of the energy-resolved conductance spectroscopy. Subsequently, the enhancement of conductance due to Andreev reflection within the energy window of the superconducting gap enables the probing of the gap under pressure. In addition to the detection of the superconducting gap, PCAR has also been used to probe chiral Majorana modes in a quantum anomalous Hall insulator~\cite{Shen2020}. Thus, PCAR shows great potential to probe exotic quantum phenomena.

To achieve a high-pressure environment, we employ the diamond anvil cell technique, which has long been a crucial workhorse in high pressure research~\cite{Eremets1996}. In a diamond anvil cell, a closed chamber filled with pressure transmitting medium is formed by pressing two opposing diamond anvils on a gasket with a central hole. Although the need to bring wires into the pressure chamber remains a challenging issue, the ``device-integrated diamond anvil cell'' (DIDAC) technique developed by some of us~\cite{Xie2021} provides a viable way of introducing the electrical contacts needed for PCAR. In particular, the small electrodes in our DIDAC device are ideal for PCAR because of the relative ease of satisfying the ballistic requirement. We note that DIDAC has already enabled a range of transport-based experiments under pressure~\cite{Zhang2022,Zhang2023,Zhang2024,wang2024,Wang2025}. Thus, the implementation of PCAR expands the functionalities of the DIDAC technique, ultimately allowing a multiprobe study of quantum materials by performing distinct types of measurements on the same sample. In this paper, we demonstrate the effectiveness of this approach by presenting the results of PCAR measurement on FeSe thin flakes using DIDAC.

\begin{figure}\resizebox{8.5cm}{!}{
\includegraphics{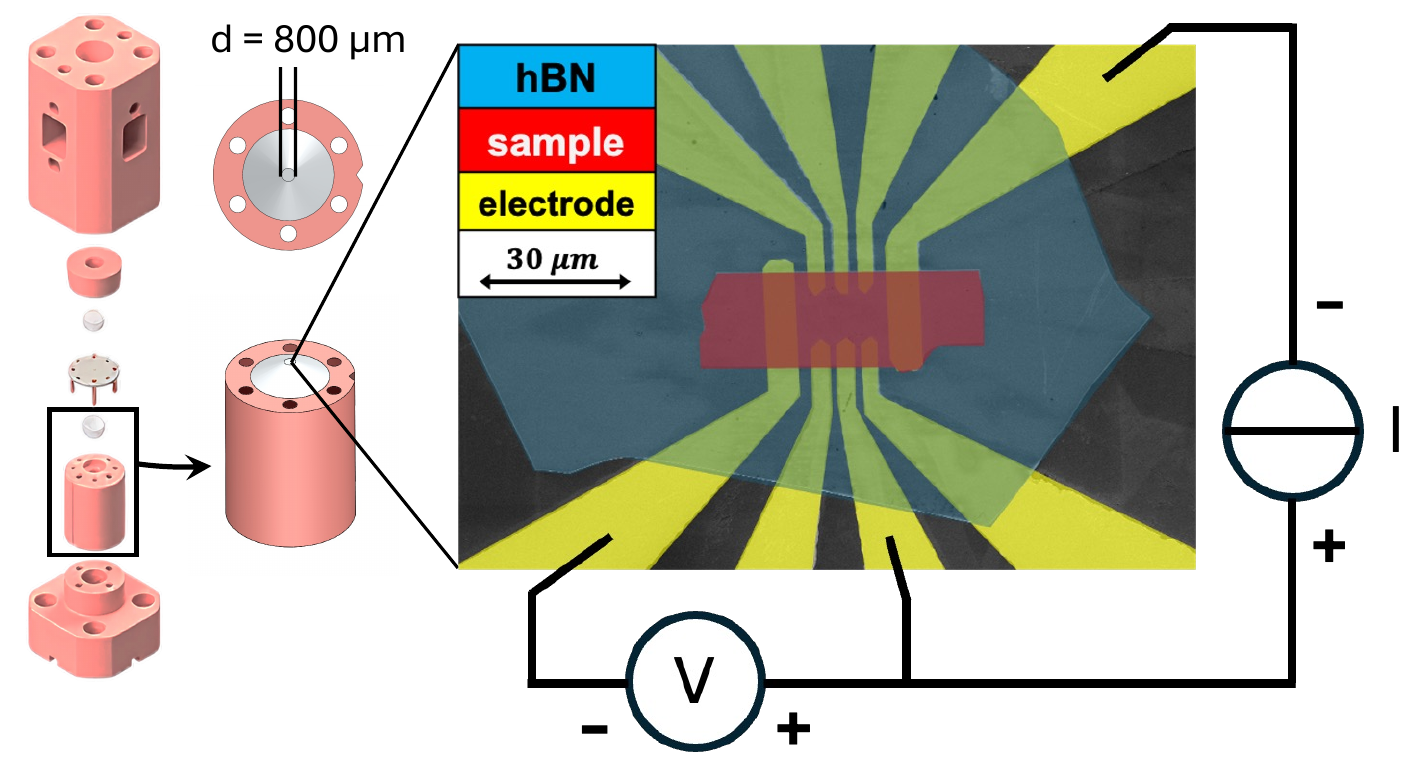}}%
\caption{\label{fig1} A schematic diagram (with false color) showing the configuration of device-integrated diamond anvil cells for point-contact spectroscopy measurement.}
\end{figure}

\section{Methodology}
Based on the concept of the ``device-integrated diamond anvil cell" technique that we previously developed\cite{Xie2021}, a diamond anvil with culet-diameter of 800 $\mu$m is lithographically patterned with 75-nm-thick gold electrodes with an intermediate layer of 5-nm-thick titanium. Two thin-flake samples, both 250--350~nm thick as determined in a focused ion beam system (Scios 2 DualBeam from Thermo Scientific), are exfoliated from the same high-quality single-crystalline FeSe grown by chemical vapor transport, and then transferred onto the diamond anvils in order to form small contacts with the gold electrodes. Hexagonal boron nitride (h-BN) flakes are used to seal the samples to avoid air exposure. One of the two thin-flake devices was prepared for the benchmark experiment at ambient-pressure, while the other one was used as one of the two anvils of the diamond anvil cell for the experiment under hydrostatic pressure (see Fig.~\ref{fig1} for schematics). The pressure transmitting medium is glycerin and the pressure achieved is measured by ruby fluorescence spectroscopy. To conduct PCAR measurements, three electrodes were selected. A DC current generated by a Keithley current source (Model 6221) was injected through the device while the bias voltage was measured by a Keysight digital multimeter (Model 34465A), as shown schematically in Fig.~\ref{fig1}. The resultant \textit{I}--\textit{V} curve was then numerically differentiated to calculate the differential conductance. The ambient-pressure benchmark experiment was conducted down to 50~mK using a Bluefors dilution refrigerator, while the high-pressure experiment with the higher $T_c$ was conducted down to 2 K and up to 14 T in a Physical Property Measurement System (PPMS) by Quantum Design.

\section{Electrical transport data and point-contact spectroscopy}
\begin{figure}\resizebox{8.5cm}{!}{
\includegraphics{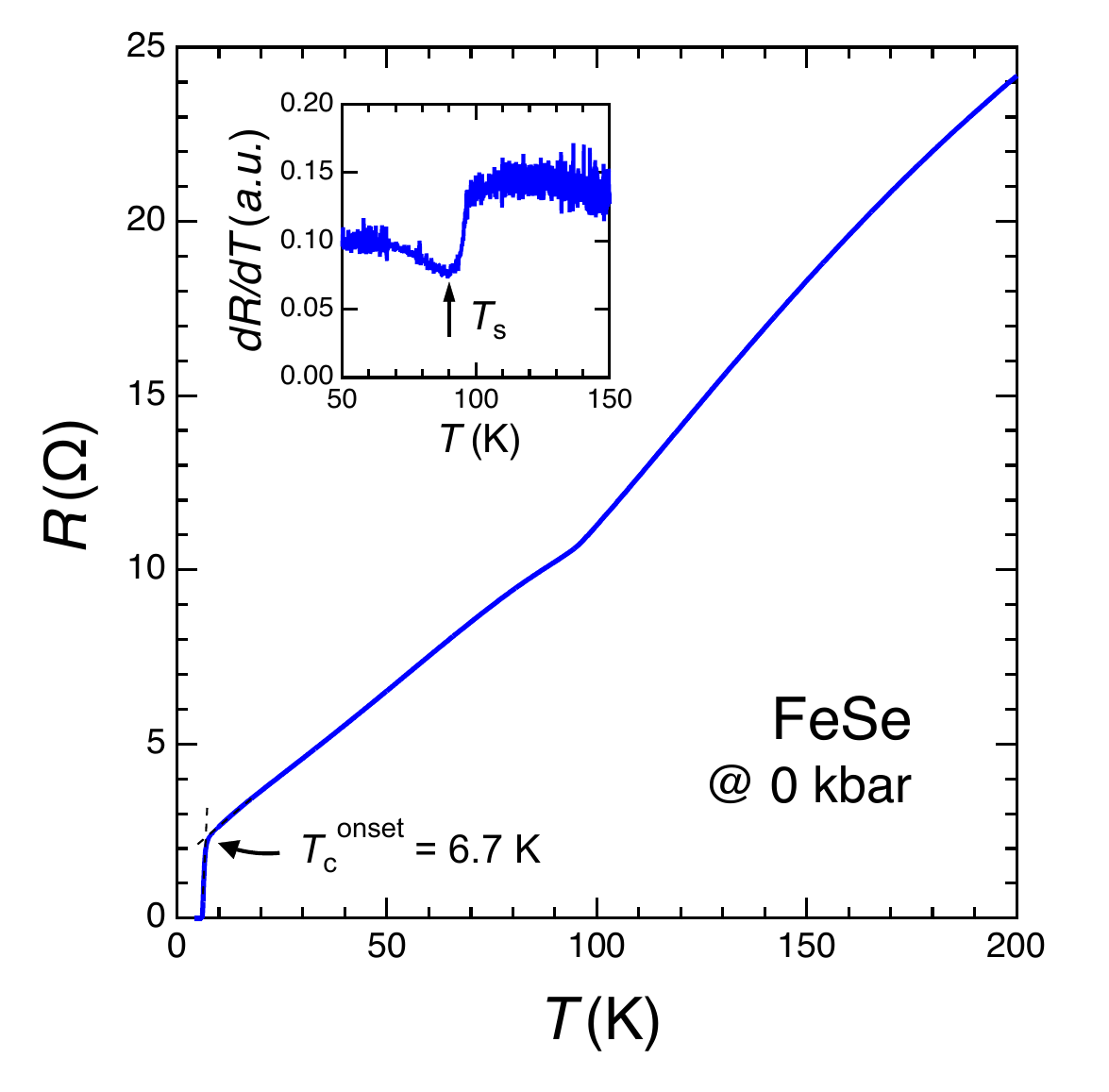}}%
\caption{\label{fig2}Temperature dependence of the electrical resistance of FeSe thin flake at ambient pressure (S1). Inset: Temperature derivative of electrical resistance for determining nematic transition temperature (\textit{T$_{\rm s}$} = 90 K).}
\end{figure}

\begin{figure}\resizebox{8.5cm}{!}{
\includegraphics{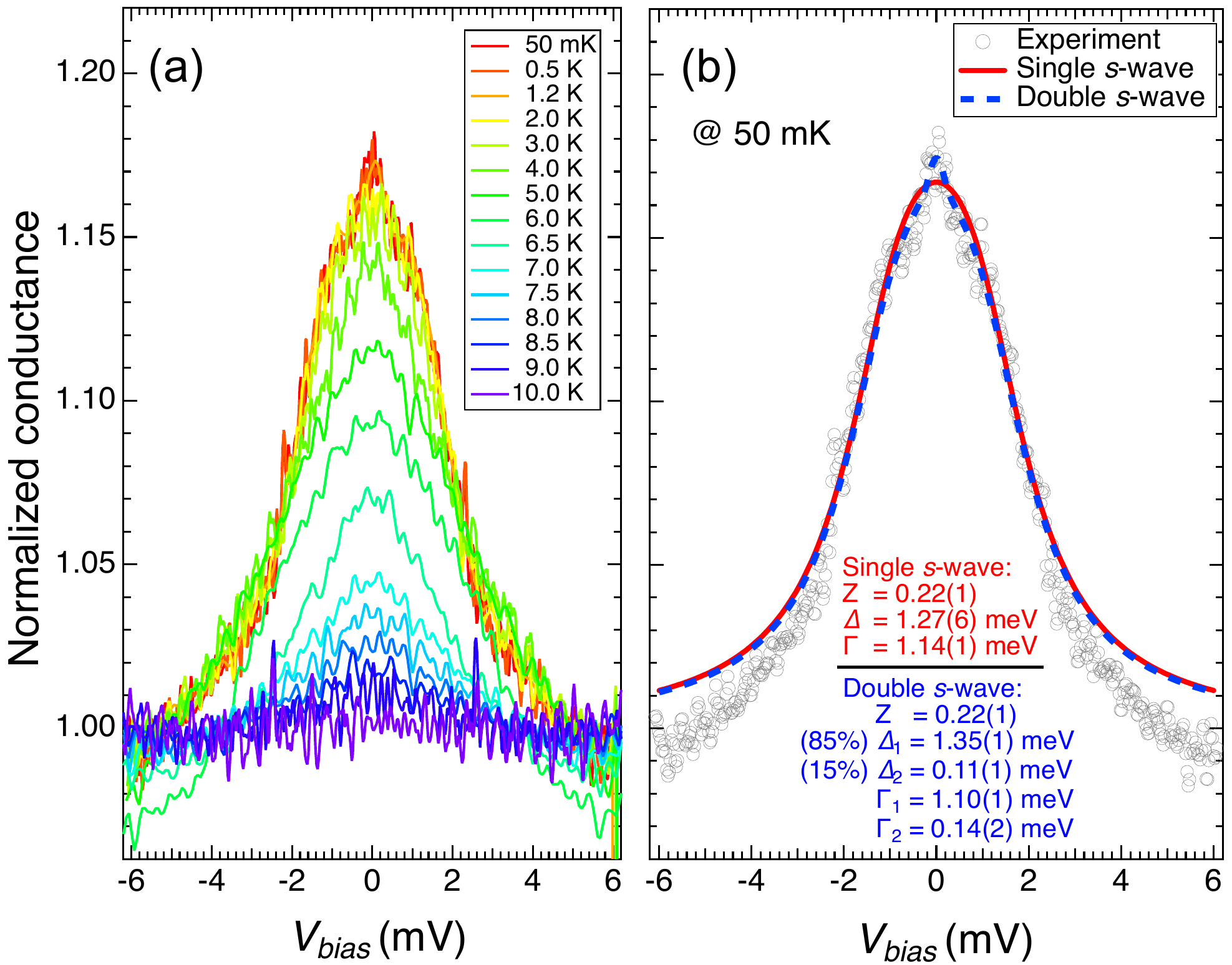}}%
\caption{\label{fig3}Normalized PCAR spectra of FeSe thin flake at ambient pressure (S1). (a) presents the temperature dependence of the normalized PCAR spectra. In (b), the normalized PCAR spectrum at 50~mK is presented by the symbols while the solid and dashed lines represent the fits by the BTK model assuming a single or double $s$-wave superconducting gap, respectively.}
\end{figure}

\subsection{Ambient-pressure benchmark}
In order to confirm the suitability of DIDAC for PCAR spectroscopy, we first test the methodology on the thin-flake device at ambient pressure. This serves to eliminate any potential complications introduced during pressurization, and enables us to establish an ambient-pressure baseline against the literature. Figure~\ref{fig2} shows the four-probe resistance measurement of the FeSe thin flake (S1) at ambient pressure. At $\sim$6.7~K, the resistance abruptly decreases to zero, indicating a superconducting transition. The nematic transition temperature ($T_{\rm s}$), determined by the anomaly in $dR/dT$ at 90~K (see the inset of Fig.~\ref{fig2}), is in close agreement with the reported values~\cite{Xie2021,Sun2016,Massat2018,Terashima2015,Li2017}.

Figure~\ref{fig3}(a) shows the normalized PCAR spectra of S1 at different temperatures. The differential conductance spectra were normalized by the background normal-state conductance (for details, see Supplementary Material). A peak near zero-bias emerges at temperature below 8~K, which is recognized as the Andreev reflection peak. The amplitude of the peak begins to saturate at around 4~K, and is stabilized down to the lowest attained temperature (50~mK). It is noteworthy that the two sharp dips on each side of the Andreev reflection peak, usually observed in the non-ballistic regime~\cite{Sheet2004, Baltz2009}, are absent in our data, indicating that our point contact is indeed in the ballistic limit.

By using the Blonder-Tinkham-Klapwijk (BTK) model~\cite{blonder1982} for single $s$-wave superconducting gap to analyze the normalized PCAR spectra (Fig.~\ref{fig3}(b)), we obtained the superconducting gap value of 1.27~meV at 50~mK, which is identical to the larger gap reported by Khasanov {\it et al.}~\cite{Khasanov2010}, where a two-gap model has been applied to describe their ambient-pressure $\mu$SR data. In fact, we have also attempted to fit our normalized PCAR spectra with a two-gap model (Fig.~\ref{fig3}(b)), which seems to give a better description of the sharp tip at zero bias that supposedly corresponds to the presence of the smaller gap. We will discuss the results of the two-gap analysis in Sec.~III B together with the high-pressure data.

\begin{figure}\resizebox{8.5cm}{!}{
\includegraphics{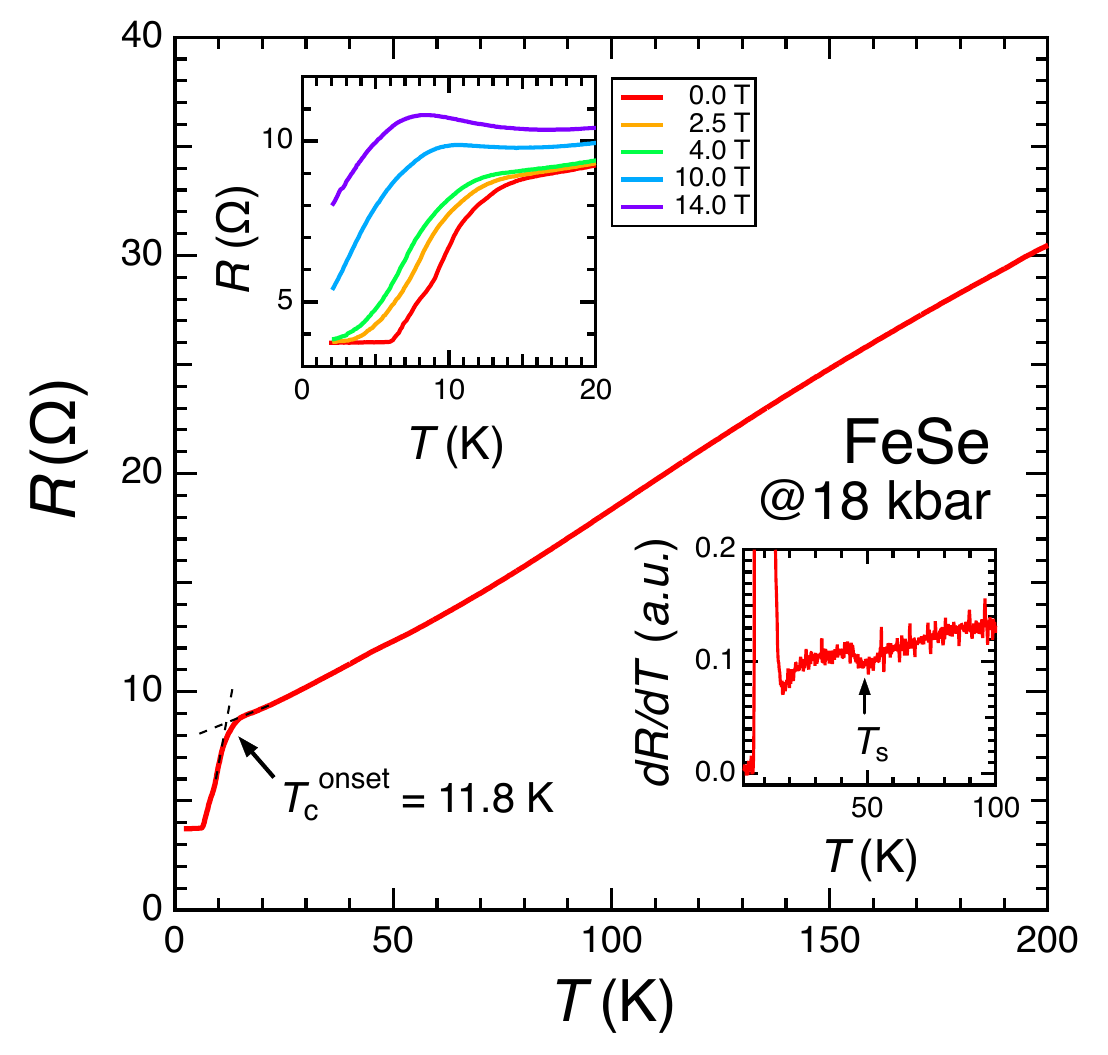}}%
\caption{\label{fig4} Temperature dependence of the electrical resistance of FeSe thin flake at 18~kbar (S2). Inset (top): Superconducting transition in electrical resistance at 18~kbar under different magnetic fields. Inset (bottom): Temperature derivative of electrical resistance, rescaled for determining nematic transition temperature (\textit{T$_{\rm s}$} = 50 K).}
\end{figure}

\begin{figure}\resizebox{8.5cm}{!}{
\includegraphics{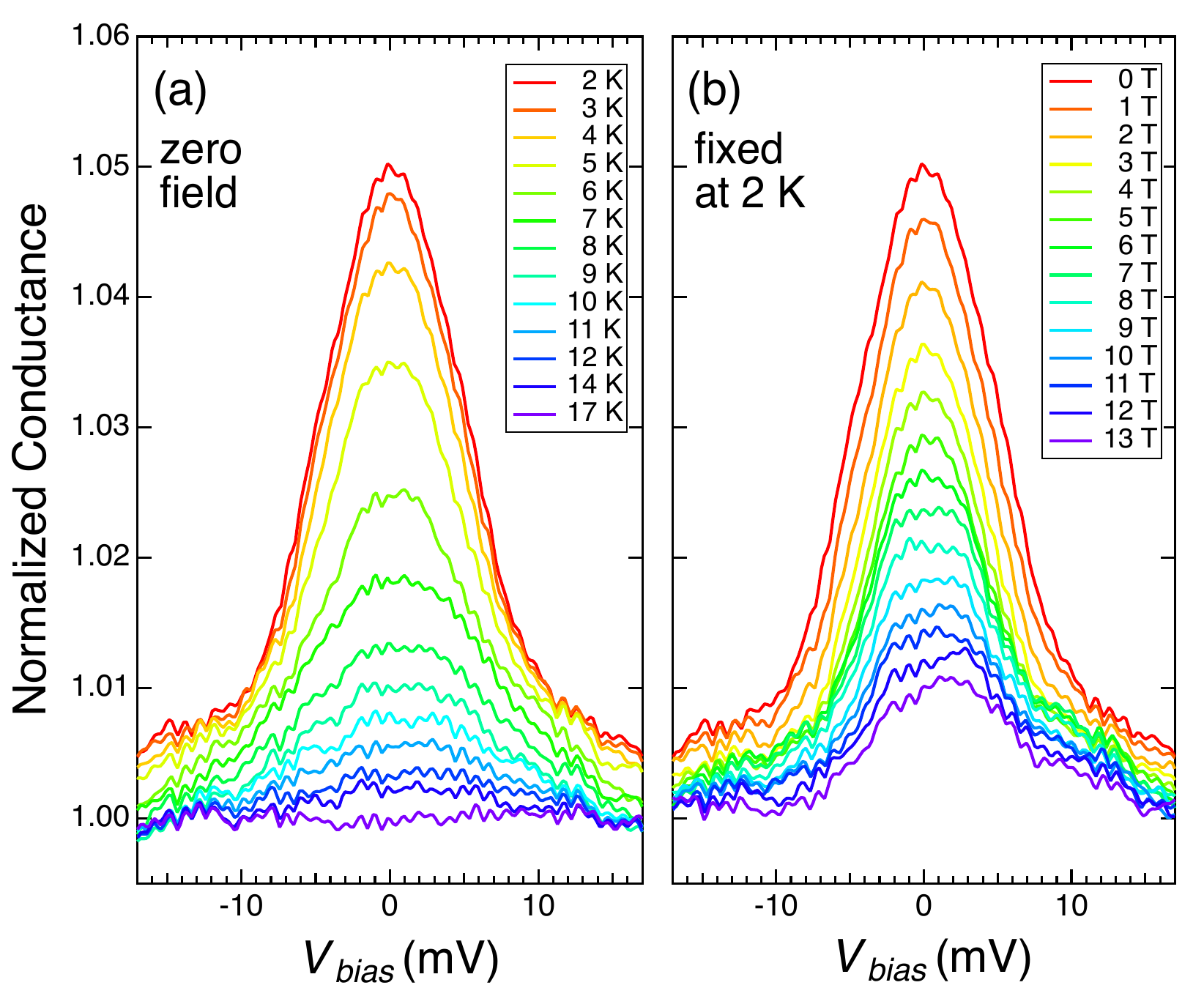}}%
\caption{\label{fig5} Normalized PCAR spectra of FeSe thin flake at 18~kbar (S2). (a) Temperature dependence at zero magnetic field. (b) Field dependence at 2 K.}
\end{figure}

\subsection{FeSe thin flake under pressure}
Encouraged by the success in observing PCAR signals at ambient pressure with the thin-flake device, we proceed to prepare another experiment under pressure, utilizing the same device architecture. Figure~\ref{fig4} shows the quasi-four-probe resistance of the second FeSe thin flake (S2) at 18~kbar. A clear and sudden drop can be detected at 11.8~K, but the resistance did not reach zero because of the usage of only three electrodes which were dedicated to PCAR configurations mentioned earlier. To show that the resistance drop indeed corresponds to a superconducting transition, we measured the resistance in a magnetic field. The resistance drop shifts to lower temperatures with an increasing field (see the top inset of Fig.~\ref{fig4}), consistent with the expected behaviour of a superconductor. Therefore, the sudden decrease of the resistance at 11.8~K is attributed to the onset transition temperature ($T_{\rm c}^{\rm ~onset}$). $T_{\rm s}$ is also discernible in the $dR/dT$ at 50~K (see the bottom inset of Fig.~\ref{fig4}), which is 40~K smaller than the ambient-pressure benchmark, indicating the suppression of the nematic phase by pressure that has been widely reported~\cite{Xie2021,Sun2016,Massat2018,Terashima2015,Li2017}.

Figure~\ref{fig5} shows the normalized PCAR spectra of S2 at different temperatures and magnetic fields. The differential conductance spectra were normalized by the parabolic background fitted with the normal-state conductance (see Supplementary Material). In the zero-field normalized conductance spectra (see Fig.~\ref{fig5}(a)), an Andreev reflection peak appears below the temperature of 14~K. The amplitude of the peak begins to saturate at around 3~K, and can be suppressed by applying an external magnetic field, as shown in Fig.~\ref{fig5}(b) for the data at 2~K. Thus, the field- and temperature-dependence of the peak reflects the behaviour of the superconductivity in the FeSe thin flake. This confirms the correspondence between the peak and the enhancement of the conductance due to the Andreev reflections within the energy range of the superconducting gap.

\begin{figure}\resizebox{8.5cm}{!}{
\includegraphics{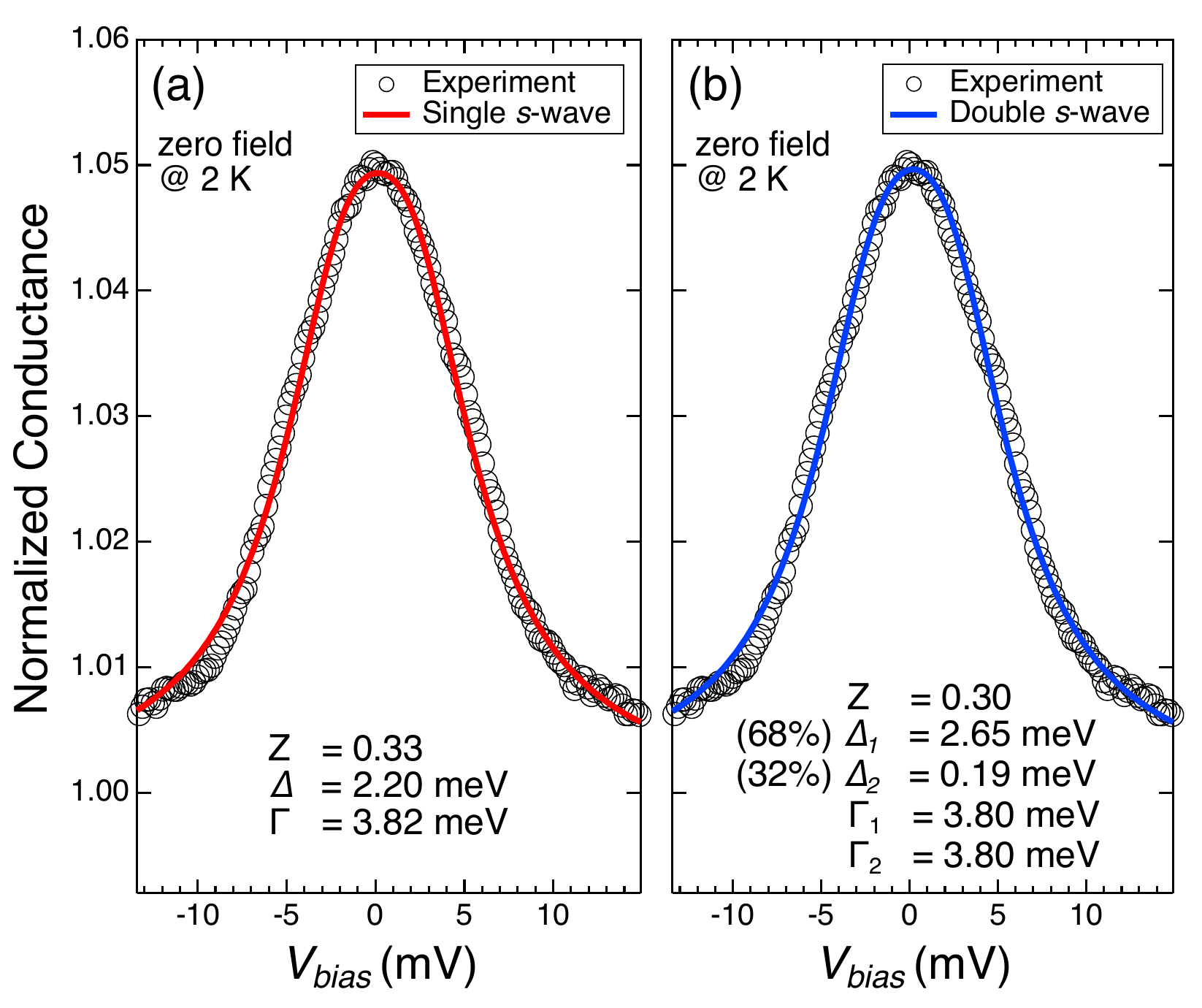}}%
\caption{\label{fig6} Normalized PCAR spectrum measured at 2~K without magnetic field (S2). The experimental data are presented by symbols, while the solid line in (a) represents a fit by the BTK model assuming a single $s$-wave superconducting gap, (b) represents a BTK fit by merging a larger $s$-wave gap with a smaller $s$-wave gap. The weight of the normalized conductance contribution for the larger gap is 68\%.}
\end{figure}

Similar to the ambient-pressure analysis, we tried to describe the normalized PCAR spectra at 18~kbar (S2) by the single $s$-wave gap and the double $s$-wave gaps BTK models. Both descriptions align well with our data at 18~kbar (S2) (see Fig.~\ref{fig6}), as they do in the case of the ambient-pressure experiment. In the single-gap scenario (Fig.~\ref{fig6}(a)), the superconducting gap obtained from the analysis is 2.20~meV, which is 1.7 times larger than the ambient-pressure baseline. Meanwhile, this value agrees well with the larger gap of 1.92~meV reported in Khasanov {\it et al.}'s two-gap analysis of their $\mu$SR data at 8.4~kbar~\cite{Khasanov2010}, the maximum pressure reached by the authors. In our attempt to fit our normalized PCAR spectra at 18~kbar (S2) with the two-gap model (Fig.~\ref{fig6}(b)), the resultant gap values are 2.65~meV and 0.19~meV, which also agree well with the literature values~\cite{Khasanov2010,Jiao2017}. However, unlike the case for the ambient-pressure benchmark, the two-gap model does not improve the fit of our PCAR spectra at 18~kbar (S2) down to 2~K, the lowest attained temperature in PPMS. Meanwhile, the model unavoidably introduces many additional fitting parameters. Furthermore, Khasanov {\it et al.}~\cite{Khasanov2010} showed that the smaller gap only opened at temperatures much lower than $T_{\rm c}$, which explains why our normalized PCAR spectra at 18~kbar (S2) do not exhibit the sharp-tip feature at zero bias. Taking into account these factors, we decide to describe all of our normalized PCAR spectra with a single-gap model, although we do not rule out the presence of two superconducting gaps suggested by other  experiments~\cite{Khasanov2010,Kasahara2014,Jiao2017,Bourgeois2016,Ponomarev2011}.Thus, our work can be interpreted as following the dominant superconducting gap from $T_c$ to the lowest attainable temperature.

\begin{figure}[!b]\resizebox{8.5cm}{!}{
\includegraphics{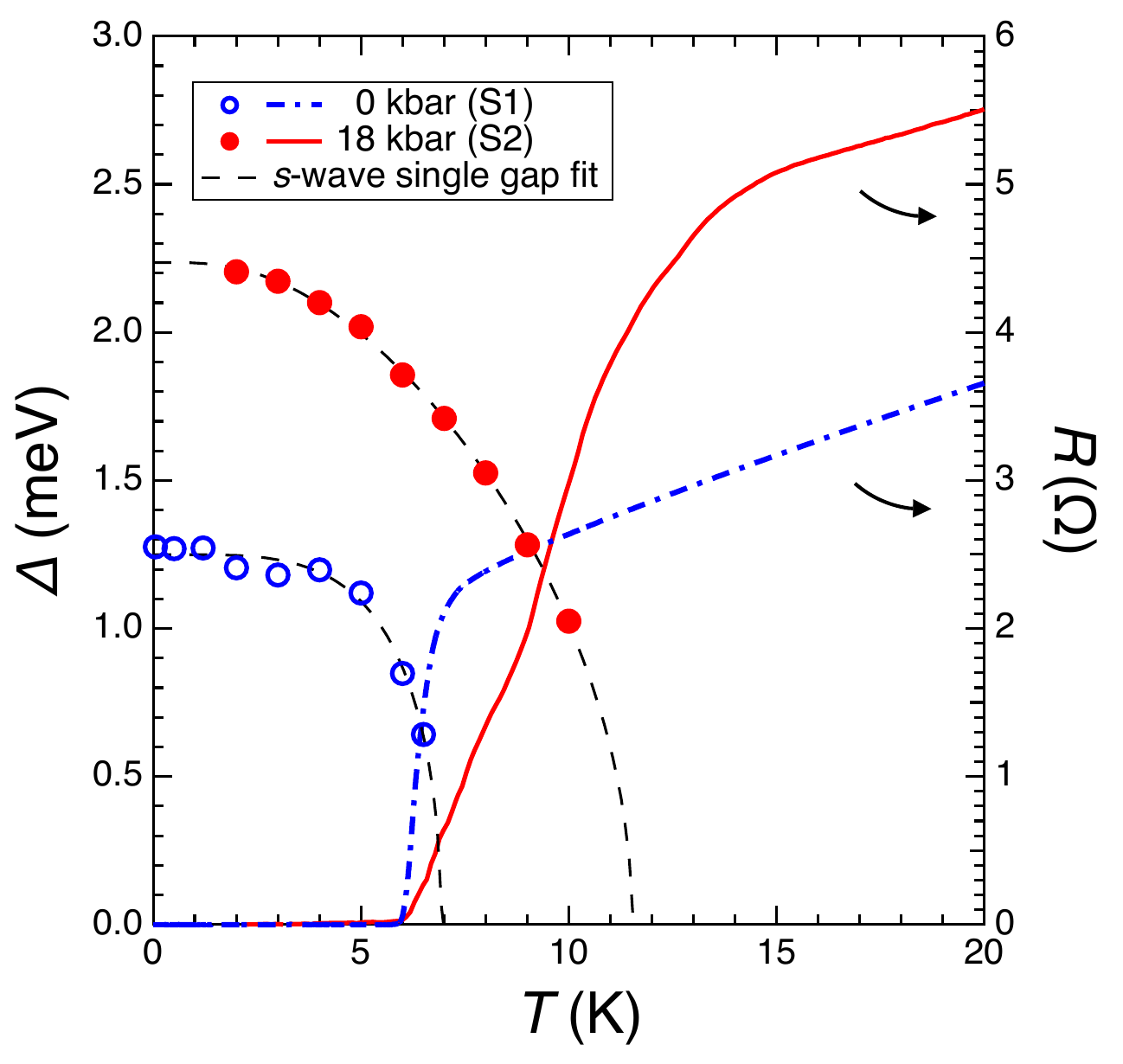}}%
\caption{\label{fig7} Temperature dependence of the superconducting gap extracted from the BTK fitting at ambient pressure (hollow symbols, S1) and 18~kbar (solid symbols, S2), and the BCS fitting results (dashed line). The temperature dependence of the electrical resistance at ambient pressure (S1) is represented by a dotted-dashed line, while the $R$--$T$ curve at 18~kbar (solid line, S2) is subtracted by the residual resistance at the lowest attainable temperature to better compare with the ambient condition.}
\end{figure}

\subsection{Temperature dependence of the dominant superconducting gap}

The temperature dependence of the superconducting gap is important because it provides crucial clue for determining the superconducting order parameter, elucidating the pairing mechanism. Figure~\ref{fig7} displays the magnitude of the dominant superconducting gap at ambient pressure (S1) and at 18~kbar (S2) over a wide temperature range. We are able to track the evolution of the gap magnitude up to 6.5~K for the ambient-pressure data, and 10~K at 18~kbar (S2), because the PCAR spectra are drastically smeared above the respective threshold values, resulting in large broadening factor ($\Gamma$). Thus, we do not view the analysis above the threshold values as conclusive. Nevertheless, with the extracted gap values shown in Fig.~\ref{fig7}, an excellent Bardeen–Cooper–Schrieffer (BCS) description of the temperature dependence of the gap magnitude can already be obtained. The resultant zero-temperature gap magnitudes are $\sim$1.25~meV at ambient pressure (S1) and $\sim$2.23~meV at 18~kbar (S2), not too different from the values discussed earlier at the lowest attainable temperatures. The fact that the temperature dependence of the gap magnitude follows the BCS description in both the ambient-pressure baseline and the case at 18~kbar lends support to our adoption of a single $s$-wave gap when we performed the BTK analysis. Finally, the BCS description shows that the gap closing temperature is $\sim$6.9~K at ambient pressure (S1) and $\sim$11.5~K at 18~kbar (S2), which are both close to the onset critical temperature determined in the resistance measurement (also shown in Fig.~\ref{fig7}). Therefore, we successfully obtained spectroscopic information of the superconducting gap in the pressurized FeSe by analyzing our PCAR data with the BTK model.

\section{Discussion}

When adapting the DIDAC to point-contact spectroscopy measurement, the small contact introduces some limitations. Since the contact area has to be as small as possible to maximize the chance of forming effective point contacts, the device is naturally vulnerable to large electrical current. However, small contact usually has large contact resistance, which allows the voltage drop in the normal metal and superconductor junction to reach a large value without flowing a large current, suggesting that large contact resistance does not forbid the observation of a complete Andreev reflection signal. Further improvement can be envisaged if multiple microelectrodes with smaller areas can be engineered. This can be explored through a laser-writing photolithography technique recently developed by some of us~\cite{Ku2022}.

Andreev reflection under pressure has also been recently observed in cerium hydrides~\cite{Cao2024} and La$_3$Ni$_2$O$_7$~\cite{Liu2025} by a similar diamond anvil cell method, but with a different way of introducing wires into the high pressure region. Our method of producing microelectrodes directly on the diamond anvil by photolithography provides another viable way of preparing ballistic contacts for detecting Andreev reflection. Meanwhile, the superconducting gaps of sulfur and H$_3$S probed by tunneling spectroscopy under pressure were recently reported by Du {\it et al.}~\cite{Du2024,Du2025}. These experiments highlight the desire to probe the superconducting gap under pressure, and our method provides a timely contribution to this important endeavour in superconductivity research. 

Our PCAR spectra exhibit a noticeable broadening. At 2~K, the broadening factor is 1.7 times greater than the resultant superconducting gap value (Fig.~\ref{fig6}(a)). According to our single-gap analysis, the broadening factor ($\Gamma$) is about 90\% of the resultant superconducting gap value at 50~mK in the ambient-pressure case, whereas the high-pressure data exhibit a $\Gamma$ value of 1.7 times the extracted gap value at 2~K. If the two-gap superconductivity scenario truly describes the system and the smaller gap is indeed an order of magnitude smaller than the dominant gap, as previously reported~\cite{Khasanov2010}, such pronounced spectral broadening can obscure the resolution of the smaller gap. This explains the absence of a clear distinction between the single- and two-gap features in our point-contact spectra under pressure (Fig.~\ref{fig6}). Spectral broadening may result from significant inelastic quasiparticle scattering at the interface between the normal metal and the superconductor. This issue can be rectified through a further reduction of the measurement temperature (particularly for S2) or by optimizing the contact quality via a more controlled fabrication environment. Another possible origin of the broadening effect could be the anisotropy of the superconducting gap, which has also been reported in FeSe by Sprau {\it et al.}~\cite{Sprau2017}. The presence of a strong anisotropy in the superconducting gap leads to a broad distribution of energy gap values around the Fermi surface, thereby contributing to spectral broadening in point-contact measurements. Nevertheless, the gap value extracted from our data agrees well with most of the reported values~\cite{Khasanov2010,Song2011,Kasahara2014,Watashige2015,Sprau2017,Jiao2017,Liu2018PRX}, suggesting that the spectral broadening does not necessarily prevent the determination of the superconducting gap value.

An interesting aspect of our data is the gap-to-$T_c$ ratio. $2\Delta(0)/k_BT_c$ calculated from our data at 18~kbar (S2) is 4.53, not only greater than the ambient-pressure baseline of 4.18 but also substantially larger than the BCS weak coupling limit. At 18~kbar, the sample is in the nematic region, but rather close to the end point of the nematicity. In the nematic phase, Raman spectroscopy has identified an anomalous softening of $A_{1g}$ Se and $B_{2g}$ Fe phonon modes~\cite{Massat2018}. If these phonon modes contribute to the pairing of electrons, the coupling strength and hence $2\Delta(0)/k_BT_c$ would be enhanced. One strength of point-contact spectroscopy is determining the gap {\it size}. Thus, the implementation of point-contact spectroscopy in a pressure cell enables us to track the gap-to-$T_c$ ratio across the nematic region in the future to explore the possible role of nematicity on the coupling strength.

In addition to the gap-to-$T_c$ ratio, the pairing mechanism of the pressure-induced superconducting state in FeSe is a rich topic. Recently, a systematic construction of temperature-pressure phase diagrams for FeSe with six different thicknesses (ranging from 5~$\mu$m to 140~nm) has been performed~\cite{Xie2021}. These phase diagrams show that FeSe exhibits extraordinary tunability by sample thickness, in addition to the well-known sensitivity to pressure. In thicker samples, $T_c$ can reach as high as 40~K under pressure~\cite{Xie2021}, but such a `high-$T_c$' phase disappears when the sample thickness is reduced. Accompanied by the disappearance of the `high-$T_c$' phase is the significant suppression of the pressure-induced magnetic region~\cite{Xie2021}. Thus, the high-pressure superconducting state of FeSe, be it thick or thin sample, deserves careful study. The DIDAC technique offers the possibility of both pressure- and thickness-tunings, and the inclusion of point-contact spectroscopy offers a comprehensive exploration of complex phases hosted by FeSe as well as in other layered superconductors.

\section{Conclusions and future prospects}
In conclusion, we have successfully functionalized our device-integrated diamond anvil cells technique to measure point-contact Andreev-reflection spectroscopy on thin flakes under pressure. The reliability of the point contact between the normal metal and the superconductor in our device is demonstrated through the measurement of point contact spectra on FeSe over a wide range of temperatures covering the superconducting transition. The temperature dependence of the gap magnitude can be well-described by the BCS theory, consistent with previous reports that show the presence of a dominant nodeless gap that appears just below the superconducting transition temperature. Since the same anvil cell design has been adopted for the measurements of quantum oscillations, the Hall effect, and critical current, the implementation of point-contact spectroscopy promises to enable a multiprobe study of thin superconducting quantum materials under pressure.

\section*{Supplementary Material}
The Supplementary Material for this paper describes the PCAR spectra background removal processes conducted in this work.

\begin{acknowledgments}
We acknowledge the financial support from the Research Grants Council of Hong Kong (GRF/14300722, GRF/14301020, and GRF/14302724) and CUHK Direct Grant (4053577, 4053664). We thank Shek Hei Chui for the useful discussion.

$^\S$C-H. K. and O. A. contributed equally to this work.
\end{acknowledgments}

\providecommand{\noopsort}[1]{}\providecommand{\singleletter}[1]{#1}%

\end{document}